\documentclass[twocolumn,showpacs,floatfix,amsmath,nofootinbib,amssymb,floatfix]{revtex4}
\usepackage{graphicx,color,dcolumn,booktabs,bm}
\usepackage{longtable,lscape}
\usepackage{txfonts}
\usepackage{overpic}
\usepackage{amssymb}
\usepackage{indentfirst}
\usepackage{feynmf}   
\usepackage{slashed}  
\usepackage{cases}
\usepackage{color}
\usepackage{multirow}
\usepackage{longtable}
\usepackage{ulem}
\usepackage{enumerate}
\usepackage{graphicx,color,dcolumn,booktabs,bm}
\usepackage[colorlinks,
            citecolor=blue,
            anchorcolor=red,
            menucolor=red,
            linkcolor=red,
            filecolor=red,
            runcolor=red,
            urlcolor=blue,
            frenchlinks=red]{hyperref}
\usepackage{epstopdf}

\begin{document}

\title{Bottomonium spectrum in the relativistic flux tube model}
\author{Bing Chen$^{1}$}\email{chenbing@ahstu.edu.cn}
\author{Ailin Zhang$^{2}$}\email{zhangal@shu.edu.cn}
\author{Jin He$^{1}$}\email{hejin@ahstu.edu.cn}
\affiliation{$^1$School of Electrical and Electronic Engineering, Anhui Science and Technology University, Fengyang 233100, China\\
$^2$Department of Physics, Shanghai University, Shanghai 200444, China}

\date{\today}

\begin{abstract}
The bottomonium spectrum is far from being established. The structures of higher vector states, including the $\Upsilon(10580)$, $\Upsilon(10860)$, and $\Upsilon(11020)$ states, are still in dispute. In addition, whether the $\Upsilon(10750)$ signal which was recently observed by the Belle Collaboration is a normal $b\bar{b}$ state or not should be examined. Faced with such a situation, we carried out a systematic investigation of the bottomonium spectrum in the scheme of the relativistic flux tube (RFT) model. A Chew-Frautschi like formula was derived analytically for the spin average mass of bottomonium states. We further incorporated the spin-dependent interactions and obtained a complete bottomonium spectrum. We found that the most established bottomonium states can be explained in the RFT scheme. The $\Upsilon(10750)$, $\Upsilon(10860)$, and $\Upsilon(11020)$ could be predominantly the $3^3D_1$, $5^3S_1$, and $4^3D_1$ states, respectively. Our predicted masses of $1F$ and $1G$ $b\bar{b}$ states are in agreement with the results given by the method of lattice QCD, which can be tested by experiments in future. We also compared the RFT model with the quark potential model in detail. The differences of these two kinds of models were discussed.

\end{abstract}
 \maketitle

\section{Introduction}

The toponium system ($t\bar{t}$) can hardly exist in nature due to the very short lifetime of the top quark ($\approx0.5\times$10$^{-24}$s)~\cite{Bigi:1986jk}. Then the bottomonium is the heaviest meson system which has been researched by experiments for many years. This fact makes the bottomonium family occupy an important position in the hadron zoo and play a special role in the study of the strong interactions. A prominent feature of the bottomonium spectrum is that many excited states are below the threshold $B\bar{B}$, which provides a good platform to test the different kinds of effective theories and phenomenological models.

Comparing with the theoretical expectations, however, the complete bottomonium spectrum is far from being established. The first three bottomonium states, namely $\Upsilon(1S)$, $\Upsilon(2S)$, and $\Upsilon(3S)$, were observed by the E288 Collaboration at Fermilab in 1977~\cite{Herb:1977ek,Innes:1977ae}. Since then nearly twenty bottomonium states have been established~\cite{Tanabashi:2018oca}. The experimental history of the $b\bar{b}$ states has been reviewed in Ref.~\cite{Segovia:2016xqb}.  Here, we just briefly review some important measurements of bottomonium in the past fifteen years.

\begin{figure}[htbp]
\begin{center}
\includegraphics[width=8.6cm,keepaspectratio]{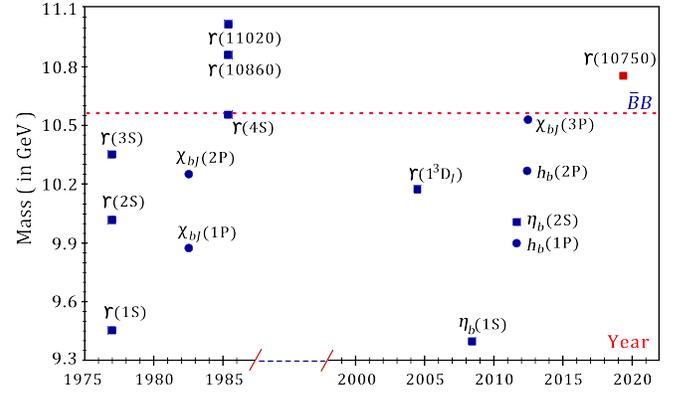}
\caption{The bottomonium states and their observed years. }\label{Fig1}
\end{center}
\end{figure}

As shown in Fig.~\ref{Fig1}, after the discovery of $\Upsilon(4S)$, $\Upsilon(10860)$, and $\Upsilon(11020)$ states~\cite{Besson:1984bd,Lovelock:1985nb}, no progress had been made in searching for the excited $b\bar{b}$ states for a long time until the CLEO Collaboration observed a $1^3D_2$ candidate in the cascade process, $\Upsilon(3S)\to\gamma\chi_b(2P)\to\gamma\gamma\Upsilon(1^3D_2)\to\gamma\gamma\gamma\chi_b(1P)\to\gamma\gamma\gamma\gamma\Upsilon(1S)$, in 2004~\cite{Bonvicini:2004yj}. This 1$D$ state was later confirmed by $BABAR$ through the $\Upsilon(1^3D_2)\to\pi^+\pi^-\Upsilon(1S)$ decay mode~\cite{delAmoSanchez:2010kz}. Furthermore, the $BABAR$ sample may contain the $\Upsilon(1^3D_1)$ and $\Upsilon(1^3D_3)$ events though the significances of these two states were very low~\cite{delAmoSanchez:2010kz}.

The spin singlet states of $S$- and $P$-wave $b\bar{b}$ mesons, i.e., $\eta_b(1S)$, $\eta_b(2S)$, $h_b(1P)$, and $h_b(2P)$, have also been found by experiments in recent years. As a long-sought state, the $\eta_b(1S)$ state was first observed by $BABAR$ in the decay channel $\Upsilon(3S)\to\gamma\eta_b(1S)$~\cite{Aubert:2008ba}, and subsequently confirmed in the decay channel $\Upsilon(2S)\to\gamma\eta_b(1S)$~\cite{Aubert:2009as}. The $\eta_b(1S)$ has also been observed by the CLEO Collaboration in the channel $\Upsilon(3S)\to\gamma\eta_b(1S)$~\cite{Bonvicini:2009hs}, and by the Belle Collaboration in the channels $h_b(nP)\to\gamma\eta_b(1S)$ ($n=1$ and 2)~\cite{Mizuk:2012pb,Tamponi:2015xzb}.

The first probable signal of the $\eta_b(2S)$ state was detected by the $BABAR$ Collaboration~\cite{Lees:2011mx} although their result was largely inconclusive. A clear evidence of $\eta_b(2S)$ was achieved by the Belle Collaboration in the processes $e^+e^-\to\Upsilon(5S)\to{h_b}(2P)\pi^+\pi^-\to\gamma\eta_b(2S)\pi^+\pi^-$~\cite{Mizuk:2012pb}. There the mass of $\eta_b(2S)$ was measured by Belle as $9999.0\pm3.5^{+2.8}_{-1.9}$ MeV.\footnote{Dobbs $et~al$. analyzed $(9.32\pm0.19)\times10^6$~$\Upsilon(2S)$ recorded with the CLEO III detector and announced the observation of $\eta_b(2S)$ in the reaction $\Upsilon(2S)\to\gamma\eta_b(2S)$~\cite{Dobbs:2012zn}. However, their result was not confirmed by Belle with a larger sample of $\Upsilon(2S)$ decays~\cite{Sandilya:2013rhy}.}


The first evidence of spin-singlet state $h_b(1P)$ was reported by $BABAR$ in the sequential decays $\Upsilon(3S)\to\pi^0h_b(1P)\to\pi^0\gamma\eta_b(1S)$~\cite{Lees:2011zp}. There the mass value of $h_b(1P)$ was measured as $9902\pm4\pm1$ MeV though the effective signal significance was only 3.0 $\sigma$. The significant signal of $h_b(1P)$ was achieved by Belle~\cite{Mizuk:2012pb,Adachi:2011ji} in the $\pi^+\pi^-$ missing spectrum of the reaction $e^+e^-\to\Upsilon(5S)\to{h_b(1P)\pi^+\pi^-}$. Meanwhile, the radial excited $h_b(2P)$ was also observed in these measurements.\footnote{Belle also measured $R\equiv\frac{\sigma(h_b(np)\pi^+\pi^-)}{\sigma(\Upsilon(2S)\pi^+\pi^-)}$ $(n=1,~2)$ and the result indicated that the $\Upsilon(5S)\to{h_b(np)}\pi^+\pi^-$ and $\Upsilon(5S)\to\Upsilon(2S)\pi^+\pi^-$ processes have similar production ratios~\cite{Adachi:2011ji}. This interesting result not only implied the complicated structure of high excited $\Upsilon$ states~\cite{Bondar:2016hva}, but also provided a new route to search the unknown $b\bar{b}$ states.} The $h_b(1P)$ state was also found in the transition $\Upsilon(4S)\to\eta{h_b(1P)}$~\cite{Tamponi:2015xzb}.

A $\chi_b(3P)$ state was first discovered by the ATLAS Collaboration in the radiative decay modes of $\chi_b(3P)\to\Upsilon(1S,2S)\gamma$~\cite{Aad:2011ih}, and subsequently confirmed by the D0~\cite{Abazov:2012gh} and LHCb Collaborations~\cite{Aaij:2014caa,Aaij:2014hla}. However, their measured masses were a little different from each other (see Table~\ref{table1}).

\begin{table}[t]
\caption{The measured mass and the observed decay mode for the $\chi_{b(J)}(3P)$ state by the different collaborations.
}\label{table1}
\renewcommand\arraystretch{1.2}
\begin{tabular*}{85mm}{c@{\extracolsep{\fill}}ccl}
\toprule[1pt]\toprule[1pt]
 State           &  Mass (MeV)              &  Decay mode          & Collaboration \\
\midrule[1pt]
$\chi_{bJ}(3P)$      & 10530$\pm$5$\pm$9                   & $\Upsilon(1S)\gamma$, $\Upsilon(2S)\gamma$         & ATLAS~\cite{Aad:2011ih}   \\
$\chi_{bJ}(3P)$      & 10551$\pm$14$\pm$17                 & $\Upsilon(1S)\gamma$                               & D0~\cite{Abazov:2012gh}   \\
$\chi_{b1}(3P)$      & 10515.7$^{+2.2+1.5}_{-3.9-2.1}$     & $\Upsilon(1S)\gamma$, $\Upsilon(2S)\gamma$,        & LHCb~\cite{Aaij:2014caa}   \\
$\chi_{b1}(3P)$      & 10511.3$\pm$1.7$\pm$2.5             & $\Upsilon(3S)\gamma$                               & LHCb~\cite{Aaij:2014hla}   \\
\bottomrule[1pt]\bottomrule[1pt]
\end{tabular*}
\end{table}

Very recently, the Belle Collaboration discovered a new candidate of the upsilon resonance in the shape of cross sections of $e^+e^-\rightarrow\Upsilon(nS)\pi^+\pi^-$ ($n=$1, 2, 3)~\cite{Abdesselam:2019gth}. Belle denoted this state as the $\Upsilon(10750)$ and determined the mass and width as
\begin{equation}
\textrm{M}=10752.7\pm5.9^{+0.7}_{-1.1}~\textrm{MeV},~~~~
\Gamma=35.5^{+17.6+3.9}_{-11.3-3.3}~\textrm{MeV}, \label{eq1}
\end{equation}
respectively, by the Breit-Wigner parametrization. Surely, more experimental confirmations are required for the $\Upsilon(10750)$ state.

Obviously, it is not an easy task to establish the bottomonium spectrum completely because even many $b\bar{b}$ states below the $B\bar{B}$ threshold have not been discovered. However, the situation may be changed especially because of the running of Belle II~\cite{Kou:2018nap}. It is expected that more excited bottomonium states will be detected in the near future. So it is time to investigate the spectrum of $b\bar{b}$ by different approaches which incorporate the spirit of QCD.

So far, different types of quark potential model have been applied in studying the bottomonium spectrum, including the nonrelativistic~\cite{Deng:2016ktl,Li:2009nr,Shah:2012js,Segovia:2016xqb,Soni:2017wvy}, the semirelativistic~\cite{Gupta:1986xt,Badalian:2004xv}, the relativized~\cite{Godfrey:1985xj,Godfrey:2015dia,Wang:2018rjg}, and the relativistic~\cite{Ebert:2011jc,Bhat:2017lde} versions. The bottomonium spectrum has also been studied by the Bethe-Salpeter equation~\cite{Fischer:2014cfa}, the coupled channel model~\cite{vanBeveren:1982qb,Tornqvist:1984fx,Liu:2011yp,Lu:2016mbb}, the QCD sum rule~\cite{Wang:2012gj,Azizi:2017izn}, the Regge phenomenology~\cite{Li:2004gu,Gershtein:2006ng,Wei:2010zza,Bakker:2019ynk}, the lattice QCD~\cite{Daldrop:2011aa,Lewis:2012ir,Wurtz:2015mqa}, and the method of perturbative QCD~\cite{Kiyo:2013aea}.

In this work, we will explore bottomonium spectrum in the scheme of the RFT model which can be rigorously derived from the Wilson area law in QCD~\cite{Brambilla:1995px}. The investigation of $b\bar{b}$ spectrum here by the RFT model could be regraded as an extension of our previous work~\cite{Chen:2014nyo}. There we have shown that the RFT model can describe the masses of single heavy baryons well. Especially, the predicted masses of 1$D$ $\Lambda_c^+$ and $\Lambda_b^0$ states in Ref.~\cite{Chen:2014nyo} are in good agreement with the later measurements by the LHCb Collaboration~\cite{Aaij:2017vbw,Aaij:2019amv}.

The manuscript is organized as follows. The RFT model is introduced in Sec. \ref{sec2} where a spin average mass formula of the heavy quarkonia is derived. In Sec. \ref{sec3}, we test the mass formula by the well measured $b\bar{b}$ states. In Sec. \ref{sec4}, the spin-dependent interactions are incorporated and the complete bottomonium spectrum is presented. In Sec. \ref{sec5}, we give further discussions about the differences of the RFT model and the quark potential model. Finally, the paper ends with the conclusion and summary.

\section{Spin average mass formula of the heavy quarkonia in the RFT model}\label{sec2}

\begin{figure}[htbp]
\begin{center}
\includegraphics[width=4.2cm,keepaspectratio]{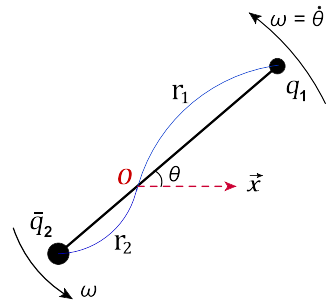}
\caption{Meson $q_1\bar{q}_2$ system in the RFT model. }\label{Fig2}
\end{center}
\end{figure}

The basic assumption of the RFT model is that the gluon field connecting the largely separated quarks in the QCD dynamical ground state could be regarded as a rigid straight tubelike color flux configuration~\cite{LaCourse:1988cu}. Thus the angular momentum of the gluon field is taken into account by the RFT model, which is qualitatively different from the usual quark potential models. The authors of Refs.~\cite{Allen:2003wz,Buisseret:2004xn} have shown that the RFT model can be derived from the Nambu-Goto QCD string model~\cite{Nambu:1969se,Susskind:1970xm,Goto:1971ce}. Furthermore, different aspects of the RFT model were investigated by different groups ~\cite{Olson:1991tw,Olsson:1992wt,Olson:1993ux,Olsson:1994dg,Olsson:1995mv,Buisseret:2004ag,Buisseret:2007hf,Buisseret:2004wm}. The deep relationship between the RFT model and QCD has also been verified in Refs.~\cite{Brambilla:1995px,Brambilla:1992pe}.  The RFT model has been applied to study the masses of heavy-light mesons~\cite{Chen:2009zt,Shan:2008ga,Jia:2018vwl}, charmonium states~\cite{Burns:2010qq}, single heavy baryons~\cite{Chen:2014nyo,Jia:2019bkr}, glueballs~\cite{Iwasaki:2003cr,Buisseret:2007de}, and other exotic hadrons~\cite{Nandan:2016uce}.

As shown in Fig.~\ref{Fig2}, the Lagrangian of a $q_1\bar{q}_2$ meson in the RFT model is written as~\cite{Iwasaki:1999py}
\begin{equation}
\mathcal{L}(r_i,\dot{\theta})=-\sum_{i=1}^{2}\left[m_i\sqrt{1-(r_i\dot{\theta})^2}+\int_0^{r_i}\tau\sqrt{1-(\rho\dot{\theta})^2}d\rho\right], \label{eq2}
\end{equation}
where $m_i$ and $r_i$ denote the mass of $i$ ($i=1$, 2) quark and its distance from the center of gravity (see Fig. \ref{Fig2}). $\tau$ represents the string (flux tube) tension. Here, we only consider the transverse velocity of the quark and antiquark, i.e., $\dot{r}_i=0$.
Then the total orbital angular momentum $L$ is defined by
\begin{equation}
L=\frac{\partial\mathcal{L}}{\partial\dot{\theta}}=\sum_{i=1}^{2}\left[\frac{m_ir_i^2\dot{\theta}}{\sqrt{1-(r_i\dot{\theta})^2}}+\int_0^{r_i}\frac{\tau\rho^2\dot{\theta}}{\sqrt{1-(\rho\dot{\theta})^2}}d\rho\right]. \label{eq3}
\end{equation}
The Hamiltonian of $q_1\bar{q}_2$ meson is given by
\begin{equation}
H=\dot{\theta}L-\mathcal{L}=\sum_{i=1}^{2}\left[\frac{m_i}{\sqrt{1-(r_i\dot{\theta})^2}}+\int_0^{r_i}\frac{\tau}{\sqrt{1-(\rho\dot{\theta})^2}}d\rho\right]. \label{eq4}
\end{equation}
When we denote the velocity of the $i$ quark which is attached with the flux tube as $u_i=r_i\dot{\theta}=r_i\omega$, the energy and orbital angular momentum can be written as
\begin{equation}
\epsilon=\sum_{i=1}^{2}\left[\frac{m_i}{\sqrt{1-u_i^2}}+\frac{\tau}{\omega}\int_0^{u_i}\frac{dv}{\sqrt{1-v^2}}\right], \label{eq5}
\end{equation}
and
\begin{equation}
L=\sum_{i=1}^{2}\left[\frac{m_iu_i^2}{\omega\sqrt{1-u_i^2}}+\frac{\tau}{\omega^2}\int_0^{u_i}\frac{v^2dv}{\sqrt{1-v^2}}\right]. \label{eq6}
\end{equation}
We have set $c=1$ in natural units for simplicity. Eqs.~(\ref{eq5}) and (\ref{eq6}) have also been obtained by the Wilson area law~\cite{Brambilla:1995px}. With Equations.~(\ref{eq5}) and (\ref{eq6}), a mass formula for the heavy-light hadrons has been derived analytically in our previous work~\cite{Chen:2014nyo}. For the bottomonium system, the masses of $b$ and $\bar{b}$ quarks are denoted as $m$. Then Eqs.~(\ref{eq5}) and (\ref{eq6}) become as
\begin{equation}
\epsilon=\frac{2m}{\sqrt{1-u^2}}+\frac{2\tau}{\omega}\arcsin u, \label{eq7}
\end{equation}
and
\begin{equation}
L=\frac{2mu^2}{\omega\sqrt{1-u^2}}+\frac{\tau}{\omega^2}\left(\arcsin u-u\sqrt{1-u^2}\right). \label{eq8}
\end{equation}
Combining with the following relationship in the RFT model,
\begin{equation}
\frac{\tau}{\omega}=\frac{mu}{1-u^2}, \label{eq9}
\end{equation}
we have
\begin{equation}
\epsilon=\frac{2m}{\sqrt{1-u^2}}+\frac{2mu}{1-u^2}\arcsin u, \label{eq10}
\end{equation}
and
\begin{equation}
\tau L=\frac{2m^2u^3}{(1-u^2)^{3/2}}+\frac{m^2u^2}{(1-u^2)^2}\left(\arcsin u-u\sqrt{1-u^2}\right). \label{eq11}
\end{equation}
Since Eqs.~(\ref{eq7}) and (\ref{eq8}) can be derived from the QCD~\cite{Brambilla:1995px}, the $m$ in the above equations could be regarded as the ``current quark masses'' of the bottom quark. In practice, the constituent quark mass is more suitable for the phenomenological analysis. To this end, we assume
\begin{equation}
m_{b}=\frac{m}{\sqrt{1-u^2}}. \label{eq12}
\end{equation}
From Eqs.~(\ref{eq7})$-$(\ref{eq11}), we have
\begin{equation}
\epsilon=2m_{b}\left(1+f_1(u)\right);~~~~\tau L=2m_b^2f_2(u). \label{eq13}
\end{equation}
In the above equations, we set the following functions,
\begin{equation}
f_1(u)=\frac{u}{\sqrt{1-u^2}}\arcsin u, \label{eq14}
\end{equation}
and
\begin{equation}
f_2(u)=\frac{u^3}{\sqrt{1-u^2}}+\frac{u^2}{2(1-u^2)}\left(\arcsin u-u\sqrt{1-u^2}\right). \label{eq15}
\end{equation}

\begin{figure}[htbp]
\begin{center}
\includegraphics[width=7.6cm,keepaspectratio]{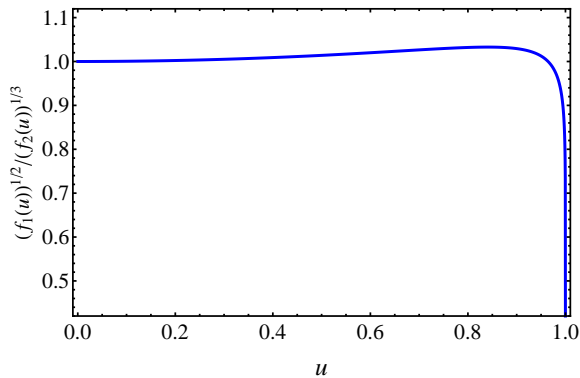}
\caption{The variation of ratio $\left(f_1(u)\right)^{1/2}/\left(f_2(u)\right)^{1/3}$ with the velocity of bottom quark $u$ in the bottomonium system. }\label{Fig3}
\end{center}
\end{figure}

Since $m_b$ has included the relativistic effect, we may treat it as the constituent quark mass of the $b$ quark. The treatment of $m_b$ which includes the relativistic effect is different from the work~\cite{Burns:2010qq} where the RFT model has been applied to investigate the assignment of $X(3872)$. As shown later, the velocity of bottom quark in the $b\bar{b}$ meson is no more than 0.50~$c$. Then Eqs.~(\ref{eq13}) can be expanded as
\begin{equation}
\begin{split}
\frac{\epsilon-2m_b}{2m_b}=&f_1(u)\approx{u^2+\mathcal{O}(u^4)+\cdots},\\
\frac{\tau L}{2m^2_b}=&f_2(u)\approx{u^3+\mathcal{O}(u^5)+\cdots}. \label{eq16}
\end{split}
\end{equation}
If we ignore the higher order of $u$, the following relationship can be obtained:
\begin{equation}
\epsilon_L=2m_b+\left(\frac{2}{m_b}\right)^{1/3}\left(\tau L\right)^{2/3}. \label{eq17}
\end{equation}
However, the validity of Eq.~(\ref{eq17}) is independent of the expansion method in Eqs.~(\ref{eq16}) since the relationship $\left(f_1(u)\right)^{1/2}\simeq\left(f_2(u)\right)^{1/3}$ always holds when the velocity of the bottom quark is taken from $0.0~c$ to $0.9~c$. To illustrate this point, the variation of ratio $\left(f_1(u)\right)^{1/2}/\left(f_2(u)\right)^{1/3}$ with the velocity of $b$ quark in the bottomonium system is presented in Fig.~\ref{Fig3}.

In the following, we replace the string tension $\tau$ by the parameter $\sigma$ with the relationship $\sigma\equiv2\pi\tau$. As done in Ref.~\cite{Chen:2014nyo}, we further extend Eq.~(\ref{eq17}) to include the radial excited $b\bar{b}$ states,
\begin{equation}
\epsilon_{nL}=2m_b+\left(\frac{\sigma^2}{2\pi^2m_b}\right)^{1/3}(\lambda n+L)^{2/3}. \label{eq18}
\end{equation}
This is a Chew-Frautschi like formula of the mass of $b\bar{b}$ states. In our calculation, the mass of $b$ quark $m_b$, the string tension parameter $\sigma$, and the dimensionless coefficient $\lambda$ in Eq.~(\ref{eq18}) are determined directly by the the well established $1S$, $1P$, and $2S$ $b\bar{b}$ states (see Sec. \ref{sec3} for details). When the distance between the $b$ and $\bar{b}$ quarks in a $b\bar{b}$ meson is denoted as $r$, we have the relationship $r=2u/\omega$. Combing with Eq.~(\ref{eq9}), we get
\begin{equation}
r=\frac{4\pi{m_b}}{\sigma}\frac{u^2}{\sqrt{1-u^2}}. \label{eq19}
\end{equation}
In the region of $u\in0.3c\sim0.6c$, we find $u^2/\sqrt{1-u^2}\approx(0.95\pm0.02)\times{f_1(u)}$. With Eqs. (\ref{eq13}) and (\ref{eq18}), we obtain the expression of $r$ as
\begin{equation}
r=\left(\frac{10.8}{\sigma{m_b}}\right)^{1/3}(\lambda n+L)^{2/3}. \label{eq20}
\end{equation}
In next Section, we test the Eq.~(\ref{eq18}) by the measured masses of $b\bar{b}$ states. In Sec. \ref{sec4}, we will incorporate the spin-dependent interactions and present a complete bottomonium spectrum.

\section{Testing Eq.~(\ref{eq18}) by the measured masses of bottomonium states} \label{sec3}

Three parameters in Eq.~(\ref{eq18}), namely the mass of bottom quark $m_b$, the string tension $\sigma$, and the dimensionless $\lambda$, should be fixed by the experimental data. We used the spin average masses of the $1S$, $2S$, and $1P$ $b\bar{b}$ states to fix the $m_b$, $\sigma$, and $\lambda$. The spin average mass of $1S$ $b\bar{b}$ is
\begin{equation}
\bar{\textrm{M}}(1S)=\frac{9398.7+9460.3\times3}{4}=9444.9 \textrm{MeV}, \label{eq21}
\end{equation}
and the average mass of $2S$ $b\bar{b}$ is
\begin{equation}
\bar{\textrm{M}}(2S)=\frac{9999+10023.3\times3}{4}=10017.2 \textrm{MeV}. \label{eq22}
\end{equation}
Here, the masses of 1$S$ and 2$S$ $b\bar{b}$ states are taken from the latest ``review of particle physics'' (RPP)~\cite{Tanabashi:2018oca} by the Particle Data Group (PDG). Since the average mass of $1^3P_0$, $1^3P_1$, and $1^3P_2$ $b\bar{b}$ states is quite close to the $1^1P_1$ state (see Ref.~\cite{Burns:2011fu} for more discussions), we take the mass of $h_b(1P)$ as the average mass of $1P$ $b\bar{b}$ states. Specifically, the world average mass of $h_b(1P)$ state, i.e., 9899.3 MeV~\cite{Tanabashi:2018oca}, is used to fix the parameters in Eq.~(\ref{eq18}). With the masses of $\bar{\textrm{M}}(1S)$, $\bar{\textrm{M}}(2S)$, and $h_b(1P)$, the parameters are fixed as
\begin{equation}
m_b=4.7224~\textrm{GeV},~~~~~\sigma=2.96~\textrm{GeV}^2,~~~~~\lambda=1.41. \label{eq23}
\end{equation}
Here, the mass of the $b$ quark which was fixed by Eq.~(\ref{eq18}) is larger than its one-loop pole mass, i.e., $m_{b,\textrm{1-loop}}$=4.550 GeV~\cite{Bakker:2019ynk}. Furthermore, the velocity of the $b$ quark could be estimated to be $0.46\pm0.01~c$ by comparing the value of $m_b$ with the current mass of the $b$ quark, i.e., $4.18^{+0.03}_{-0.02}$ GeV~\cite{Tanabashi:2018oca}.

With the values of $m_b$, $\sigma$, and $\lambda$, the center of gravity of the other $n^{2S+1}L_J$ multiplet can be calculated directly. At present, the masses of $\Upsilon(3S)$, $h_b(2P)$ and $\Upsilon_2(1D)$ states have been well measured by different experiments~\cite{Tanabashi:2018oca}. A comparison of the masses of these $b\bar{b}$ states with the predictions by Eq.~(\ref{eq18}) is given in the Table \ref{table2}.

\begin{table}[htbp]
\caption{The predicted spin average masses of the $1D$, $2P$, and $3S$ $b\bar{b}$ multiplets (in MeV). The measured masses of observed candidates~\cite{Tanabashi:2018oca} are also listed for comparison.
}\label{table2}
\renewcommand\arraystretch{1.3}
\begin{tabular*}{85mm}{c@{\extracolsep{\fill}}ccc}
\toprule[1pt]\toprule[1pt]
 $nL$           &  State              &  Measured mass          & Prediction \\
\midrule[1pt]
 $1D$           &  $\Upsilon_2(1D)$   & 10163.7$\pm$1.4              & 10166   \\
 $2P$           &  $h_b(2P)$          & 10259.8$\pm$1.2              & 10262   \\
 $3S$           &  $\Upsilon(3S)$     & 10355.2$\pm$0.5              & 10352   \\
\bottomrule[1pt]\bottomrule[1pt]
\end{tabular*}
\end{table}
The mass of $h_b(2P)$ is predicted to be 10262 MeV which is consistent with the experimental result. The $\eta_b(3S)$ state has not been discovered by the experiment. Nevertheless, the spin average mass of the $2S$ bottomonium states is about 6 MeV below the $\Upsilon(2S)$ state [see Eq.~(\ref{eq22})]. So one could reasonably expect the average mass of $3S$ states to be about 10350 MeV, which is also close to our prediction. As argued in Ref.~\cite{Lees:2011mx}, two $D$-wave $b\bar{b}$ states, namely the $\Upsilon(10152)$ and $\Upsilon_3(10173)$, may have been detected in the experimental data by the CLEO~\cite{Bonvicini:2004yj} and $BABAR$~\cite{delAmoSanchez:2010kz} collaborations. Although the measured masses of these two states need more confirmation, the average mass of the $\Upsilon(10152)$, $\Upsilon_2(10164)$, and $\Upsilon_3(10173)$ states
\begin{equation}
\frac{10152\times3+10163.7\times5+10173\times7}{15}=10165.7~\textrm{MeV} \label{eq24}
\end{equation}
is quite consistent with our result (see Table~\ref{table2}).

As shown above, the predicted average masses of $\Upsilon(3S)$, $h_b(2P)$, and $\Upsilon_2(1D)$ multiplets are well comparable with the experimental results. For completeness, we incorporate the spin-dependent interactions and give a whole bottomonium spectrum in the next section.

\section{The complete bottomonium spectrum by incorporating the spin-dependent interactions} \label{sec4}
For simplicity, we consider the color hyperfine interaction
\begin{equation}
H_{\textrm{hyp}}=\frac{4\alpha_s}{3m_b^2}\left(\frac{8\pi}{3}\delta^3(\textbf{r})\textbf{s}_b\cdot\textbf{s}_{\bar{b}}+\frac{1}{r^3}\hat{S}_{b\bar{b}}\right), \label{eq25}
\end{equation}
which arises from the one gluon exchange (OGE) forces, and the following spin-orbit term
\begin{equation}
H_{\textrm{so}}=\frac{1}{m_b^2}\left(\frac{2\alpha_s}{r^3}-\frac{b}{2r}\right)\textbf{S}\cdot\textbf{L}, \label{eq26}
\end{equation}
which includes the OGE spin orbit and the longer-ranged inverted spin-orbit terms. This type of spin-dependent interactions has been applied to investigate the mass spectrum of charmonia states~\cite{Barnes:2005pb}. The $\hat{S}_{b\bar{b}}$ denotes the tensor operator. The ``$\delta^3(\textbf{r})$'' function which comes from a contact hyperfine interaction can be simulated by different forms of smearing functions~\cite{Godfrey:1985xj,Vijande:2004he}. In our calculations, we take the following smearing function,
\begin{equation}
f(\textbf{r})=\frac{4}{\pi^2 r_0^2}\frac{e^{-\sqrt{r/r_0}}}{r}, \label{eq27}
\end{equation}
to reproduce the mass splitting of $nS$ ($n\geq2$).~\footnote{The distance of $b$ and $\bar{b}$ quarks in the $1S$ state is given as 0 by Eq.~(\ref{eq20}), which is obviously underestimated. So we do not reproduce the mass splitting of the $1^3S_1$ and $1^1S_0$ $b\bar{b}$ states.} Here, we take the $r_0$ as 0.94 GeV$^{-1}$. Because of the heavy masses, the distance between $b$ and $\bar{b}$ quarks in the low-lying bottomonium states is much smaller. Therefore, one should treat the running coupling constant $\alpha_s$ in Eqs.~(\ref{eq25}) and (\ref{eq26}) seriously. We use the following,
\begin{equation}
\alpha_s(r)=\alpha_0\textrm{Erf}\left[\left(\frac{m_b r}{0.72\pi^2}\right)^{5/2}\right], \label{eq28}
\end{equation}
to simulate the running coupling constant, where the $\textrm{Erf}[\cdots]$ refers to the error function. In our calculations, the running coupling constant is assumed to saturate at  0.68, i.e., $\alpha_0=0.68$. To reduce the free parameters, we take the value of $b$ in Eq.~(\ref{eq26}) as the string tension $\tau$ in the RFT model, i.e., $b=\sigma/2\pi=0.471$ GeV$^2$. With Eqs.~(\ref{eq20}), (\ref{eq25}), (\ref{eq27}), and (\ref{eq28}), the splitting masses of $n^3S_1$ and $n^1S_0$ states $(n\geq2)$ are presented in Table~\ref{table3}.

\begin{table}[htbp]
\caption{The mass splitting of $n^3S_1$ and $n^1S_0$ states (in MeV).
}\label{table3}
\renewcommand\arraystretch{1.2}
\begin{tabular*}{85mm}{c@{\extracolsep{\fill}}ccccc}
\toprule[1pt]\toprule[1pt]
 $\Delta\textrm{M}(nS)$    &   $n=2$   &   $n=3$     &  $n=4$  &  $n=5$  &  $n=6$   \\
\midrule[1pt]
 Our                         &   23.9    &   20.7      &  13.2   &  9.3    &  7.0   \\
 Ref.~\cite{Godfrey:2015dia} &   27      &   18        &  12     &  9      &  5    \\
 Ref.~\cite{Deng:2016ktl}    &   25      &   17        &  13     &  11     &  9   \\
\bottomrule[1pt]\bottomrule[1pt]
\end{tabular*}
\end{table}

Obviously, our results in Table~\ref{table3} are comparable with those from Refs.~\cite{Godfrey:2015dia,Deng:2016ktl}. As shown later, the masses of most known $b\bar{b}$ states can also be reproduced, though our method is quite phenomenological.

\subsection{$nS$ $(n\geq2)$ states} \label{A}

With the predicted splitting masses in Table~\ref{table3}, the masses of $n^1S_0$ and $n^3S_1$ bottomonium states ($n\geq2$) are predicted in Table~\ref{table4} where the experimental data~\cite{Tanabashi:2018oca} and the results from other works~\cite{Soni:2017wvy,Godfrey:2015dia,Liu:2011yp} are also listed for comparison.

\begin{table}[htbp]
\caption{The masses of the $nS$ ($n\geq2$) $b\bar{b}$ states (in MeV).
}\label{table4}
\renewcommand\arraystretch{1.2}
\begin{tabular*}{85mm}{c@{\extracolsep{\fill}}ccccc}
\toprule[1pt]\toprule[1pt]
 States        &  Expt.~\cite{Tanabashi:2018oca}  &   Our     &  Ref.~\cite{Godfrey:2015dia}   &  Ref.~\cite{Soni:2017wvy}  &  Ref.~\cite{Liu:2011yp}   \\
\midrule[1pt]
 $0^{-+}(2S)$  &  9999$\pm$4                &   9999    &  9976   &  9955    &  10005   \\
 $1^{--}(2S)$  &  10023.3$\pm$0.3           &   10023   &  10003  &  9979    &  10026   \\
 $0^{-+}(3S)$  &                            &   10337   &  10336  &  10338   &  10338   \\
 $1^{--}(3S)$  &  10355.2$\pm$0.5           &   10357   &  10354  &  10359   &  10352   \\
 $0^{-+}(4S)$  &                            &   10627   &  10623  &  10663   &  10593   \\
 $1^{--}(4S)$  &  10579.4$\pm$1.2           &   10637   &  10635  &  10683   &  10603   \\
 $0^{-+}(5S)$  &                            &   10878   &  10869  &  10956   &  10813   \\
 $1^{--}(5S)$  &  10889.9$^{+3.2}_{-2.6}$   &   10887   &  10878  &  10975   &  10820   \\
 $0^{-+}(6S)$  &                            &   11111   &  11097  &  11226   &  11008   \\
 $1^{--}(6S)$  &  10992.9$^{+10.0}_{-3.1}$  &   11118   &  11102  &  11243   &  11023   \\
\bottomrule[1pt]\bottomrule[1pt]
\end{tabular*}
\end{table}
As shown in Table~\ref{table4}, the masses of well measured $\eta_b(2S)$, $\Upsilon(2S)$, and $\Upsilon(3S)$ states are reproduced in our scheme. The predicted mass of the unknown $\eta_b(3S)$ state is 10337 MeV which is comparable with these results from Refs.~\cite{Godfrey:2015dia,Soni:2017wvy,Liu:2011yp}.

The masses of the $\Upsilon(4S)$, $\Upsilon(5S)$, and $\Upsilon(6S)$ obtained by the RFT model are quite close to the results given by the Godfrey-Isgur model~\cite{Godfrey:2015dia}. Our results favor the $\Upsilon(10860)$ as a predominantly $5^3S_1$ state. Interestingly, a recent work based on the lattice QCD also suggested the $\Upsilon(10860)$ as a $5^3S_1$ state~\cite{Bicudo:2019ymo}. The mass of $\Upsilon(4S)$ predicted by the RFT model is about 60 MeV higher than the measured mass of $\Upsilon(10580)$ (see Table \ref{table4}). The mass of the $\Upsilon(4S)$ state predicted in Refs.~\cite{Deng:2016ktl,Li:2009nr,Godfrey:2015dia,Segovia:2016xqb,Soni:2017wvy,Liu:2011yp} was also larger than the $\Upsilon(10580)$ state. In the quark potential models, the mass gap between the $3^3S_1$ and $4^3S_1$ $b\bar{b}$ states is expected to be larger than the gap between the $4^3S_1$ and $5^3S_1$ states. However, the experimental measurement is contrary to the expectation, i.e.,
\begin{equation}
\Delta\textrm{M}(\Upsilon(10580)-\Upsilon(10355))\approx224.2~\textrm{MeV}, \label{eq29}
\end{equation}
which is smaller than
\begin{equation}
\Delta\textrm{M}(\Upsilon(10860)-\Upsilon(10580))\approx310.5~\textrm{MeV}. \label{eq30}
\end{equation}
It indicates that the mass of $\Upsilon(4S)$ shifts down about 40$\sim$50 MeV due to a particular mechanism. This anomalously mass gap of ``$\Upsilon(4S)-\Upsilon(3S)$'' and ``$\Upsilon(5S)-\Upsilon(4S)$'' cannot simply be solved by the n\"{a}ive quark model. T\"{o}rnqvist proposed a solution to this puzzle. Specifically, it may be disentangled by considering the coupled-channel effects~\cite{Tornqvist:1984fx}. More importantly, the masses of $\Upsilon(5S)$ and $\Upsilon(6S)$ were well predicted in the scheme of the coupled-channel model~\cite{Tornqvist:1984fx} before the observations of candidates, i.e., $\Upsilon(10860)$ and $\Upsilon(11020)$~\cite{Besson:1984bd,Lovelock:1985nb}. The scheme suggested by T\"{o}rnqvist was supported by the recent work~\cite{Lu:2016mbb}.

If the $\Upsilon(11020)$ is a pure $6^3S_1$ $b\bar{b}$ state, its measured mass is about 100$\sim$200 MeV lower than the prediction by the RFT model and other methods~\cite{Godfrey:2015dia,Soni:2017wvy,Liu:2011yp,Badalian:2008ik}. So it seems that the $\Upsilon(11020)$ is not a pure $6S$ upsilon resonance. This conclusion is partially supported by the analysis of its dielectron widths~\cite{Badalian:2008ik} (see Sec. \ref{C}).

\subsection{$nP$ states} \label{B}

The masses of $nP$ ($n=1\sim5$) $b\bar{b}$ states which are predicted by the RFT model are listed in Table~\ref{table5} with the experimental data~\cite{Tanabashi:2018oca} and other theoretical results from Refs.~\cite{Godfrey:2015dia,Ebert:2011jc,Soni:2017wvy}. Up to now, the $1P$ and $2P$ bottomonium states have been well established~\cite{Tanabashi:2018oca}. Obviously, the masses of these states are well reproduced by the RFT model.

\begin{table}[htbp]
\caption{The masses of the $nP$ $b\bar{b}$ states (in MeV).
}\label{table5}
\renewcommand\arraystretch{1.2}
\begin{tabular*}{85mm}{c@{\extracolsep{\fill}}ccccc}
\toprule[1pt]\toprule[1pt]
 States        &  Expt.~\cite{Tanabashi:2018oca}  &   Our     &  Ref.~\cite{Godfrey:2015dia}   &  Ref.~\cite{Ebert:2011jc}  &  Ref.~\cite{Soni:2017wvy}   \\
\midrule[1pt]
 $0^{++}(1P)$  &  9859.4$\pm$0.7            &   9854    &  9847   &  9859    &  9806   \\
 $1^{++}(1P)$  &  9892.8$\pm$0.6            &   9893    &  9876   &  9892    &  9819   \\
 $1^{+-}(1P)$  &  9899.3$\pm$0.8            &   9899    &  9882   &  9900    &  9821   \\
 $2^{++}(1P)$  &  9912.2$\pm$0.6            &   9911    &  9897   &  9912    &  9825   \\
 $0^{++}(2P)$  &  10232.5$\pm$0.9           &   10239   &  10226  &  10233   &  10205   \\
 $1^{++}(2P)$  &  10255.5$\pm$0.7           &   10259   &  10246  &  10255   &  10217   \\
 $1^{+-}(2P)$  &  10259.8$\pm$1.2           &   10262   &  10250  &  10260   &  10220   \\
 $2^{++}(2P)$  &  10268.7$\pm$0.7           &   10268   &  10261  &  10268   &  10224   \\
 $0^{++}(3P)$  &                            &   10551   &  10522  &  10521   &  10540   \\
 $1^{++}(3P)$  &  10513.4$\pm$0.7           &   10557   &  10538  &  10541   &  10553   \\
 $1^{+-}(3P)$  &                            &   10556   &  10541  &  10544   &  10556   \\
 $2^{++}(3P)$  &  10524.0$\pm$0.8           &   10556   &  10550  &  10550   &  10560   \\
 $0^{++}(4P)$  &                            &   10815   &  10775  &  10781   &  10840   \\
 $1^{++}(4P)$  &                            &   10817   &  10788  &  10802   &  10853   \\
 $1^{+-}(4P)$  &                            &   10815   &  10790  &  10804   &  10855   \\
 $2^{++}(4P)$  &                            &   10814   &  10798  &  10812   &  10860   \\
 $0^{++}(5P)$  &                            &   11053   &  11004  & $\cdots$ &  11115   \\
 $1^{++}(5P)$  &                            &   11053   &  11014  & $\cdots$ &  11127   \\
 $1^{+-}(5P)$  &                            &   11051   &  11016  & $\cdots$ &  11130   \\
 $2^{++}(5P)$  &                            &   11049   &  11022  & $\cdots$ &  11135   \\
\bottomrule[1pt]\bottomrule[1pt]
\end{tabular*}
\end{table}

The candidates of $3P$ bottomonium states have been detected by the ATLAS~\cite{Aad:2011ih}, D0~\cite{Abazov:2012gh}, and LHCb~\cite{Aaij:2014caa,Aaij:2014hla} collaborations (see Table~\ref{table1}). The masses of the $\chi_{b1}(3P)$ and $\chi_{b2}(3P)$ collected by the PDG are listed in Table~\ref{table5}. The experimental results are about 20$\sim$40 MeV smaller than the theoretical results. One notices that the predicted masses $3P$ $b\bar{b}$ states are about 30$\sim$100 MeV above the thresholds of $B\bar{B}$, $B\bar{B}^\ast+B^\ast\bar{B}$, and $B^\ast\bar{B}^\ast$ decay channels. So the coupled-channel channel effect may affect the properties of $3P$ bottomonium states including their masses.\footnote{However, the practical calculations in Ref.~\cite{Ferretti:2018tco} did not support this conjecture. There the $\chi_b(3P)$ state was suggested to be the (almost) pure bottomonia.} More theoretical and experimental efforts are desirable for the $3P$ $b\bar{b}$ states in future.

The $4P$ and $5P$ bottomonium states are predicted around 10800 and 11050 MeV, respectively, which means these states locate above the open-bottom thresholds. Then the Okubo-Zweig-Iizuka allowed decays are probable for these states. In Ref.~\cite{Godfrey:2015dia}, the investigation of strong decays by the $^3P_0$ model indicated that the $\chi_{b0}(4P)$ state mainly decays through the $B\bar{B}$ and $B^\ast{\bar{B}}^\ast$ channels while the $B\bar{B}^\ast+B^\ast\bar{B}$ is the largest decay channel for the $\chi_{b1}(4P)$, $\chi_{b2}(4P)$, and $h_b(4P)$ states. Different from the $4P$ bottomonium states, the largest decay channel of $5P$ states is the $B^\ast{\bar{B}}^\ast$. The total decay widths of $4P$ and $5P$ bottomonium states were predicted to be 30$\sim$70 MeV. The decays predicted in Ref.~\cite{Wang:2018rjg} were slightly different from these results in Ref.~\cite{Godfrey:2015dia}. Of course, discovery of these high $P$-wave bottomonium states is a great challenge for present experiments.

\subsection{$nD$ states} \label{C}

So far only one $D$-wave $b\bar{b}$ state, namely $\Upsilon_2(1D)$, was listed in the summary table of PDG~\cite{Tanabashi:2018oca}. Its measured mass, i.e., 10163.7$\pm$1.7 MeV, is quite in agreement with our prediction (see Table~\ref{table6}). The visible evidence of the $1^3D_1$ and $1^3D_3$ bottomonium states at 10152 and 10173 MeV~\cite{Bonvicini:2004yj,delAmoSanchez:2010kz}, respectively, was pointed out in Ref.~\cite{Lees:2011mx}. Our predictions in Table~\ref{table6} are comparable with these preliminary results. Our results are also consistent with the predicted masses of 1$D$ $b\bar{b}$ states by lattice QCD~\cite{Daldrop:2011aa}.

None of the $2D$ $b\bar{b}$ states have been announced by any experiments. Nevertheless, Beveren and Rupp found the $\Upsilon(2D)$ signal with 10.7 standard deviations~\cite{vanBeveren:2010gm} by reanalyzing the $BABAR$ data~\cite{Aubert:2008az}. There the mass of $\Upsilon(2D)$ was fitted to be $10495\pm5$ MeV, which is a bit larger than the predictions in Table~\ref{table6}.

As mentioned before, a $1^{--}$ structure $\Upsilon(10750)$ which was discovered by the Belle collaboration~\cite{Abdesselam:2019gth} is still unclear. Since the $3^3D_1$ $b\bar{b}$ state is expected to has a masse around 10740 MeV, the $\Upsilon(10750)$ could be a good $3D$ candidate. Due to the significant mixing between the $(n+1)^3S_1$ and $n^3D_1$ states $(n\geq3)$, the magnitude of dielectron widths of the mixed $\tilde{\Upsilon}(n^3D_1)$ resonances $(n=3, 4, 5)$ can increase by 2 orders~\cite{Badalian:2008ik}. For the $\tilde{\Upsilon}(3D)$ state, the dielectron width was obtained to be $0.095^{+0.028}_{-0.025}$ keV, which indicated that the predominantly $3^3D_1$ $b\bar{b}$ state can be produced in the $e^+e^-$ annihilation process with the high statistics data. Furthermore, the decay width of the $3^3D_1$ $b\bar{b}$ state was obtained as 54.1 MeV~\cite{Wang:2018rjg} which is comparable with the measurement by the Belle Collaboration~\cite{Abdesselam:2019gth} [see Eq.~(\ref{eq1})]. So the $\Upsilon(10750)$ could be predominantly a $3^3D_1$ $b\bar{b}$ state in our scheme. However, the other explanations suggested in Refs.~\cite{Wang:2019veq,Li:2019qsg} are also possible for the $\Upsilon(10750)$ state. For revealing the inner structure of $\Upsilon(10750)$, more precise measurements including the dielectron width and the branching ratios of $\Gamma(B\bar{B}):\Gamma(B\bar{B}^\ast+B^\ast\bar{B}):\Gamma(B^\ast{\bar{B}}^\ast)$ are needed in future.

\begin{table}[htbp]
\caption{The masses of the $nD$ $b\bar{b}$ states (in MeV).
}\label{table6}
\renewcommand\arraystretch{1.2}
\begin{tabular*}{85mm}{c@{\extracolsep{\fill}}ccccc}
\toprule[1pt]\toprule[1pt]
 States        &  Expt.~\cite{Tanabashi:2018oca}  &   Our      &  Ref.~\cite{Godfrey:2015dia}   &  Ref.~\cite{Ebert:2011jc}  &  Ref.~\cite{Soni:2017wvy}   \\
\midrule[1pt]
 $1^{--}(1D)$  &                            &   10136   &  10138  &  10154   &  10074   \\
 $2^{--}(1D)$  &  10163.7$\pm$1.7           &   10164   &  10147  &  10161   &  10075   \\
 $2^{-+}(1D)$  &                            &   10167   &  10148  &  10163   &  10074   \\
 $3^{--}(1D)$  &                            &   10183   &  10155  &  10166   &  10073   \\
 $1^{--}(2D)$  &                            &   10467   &  10441  &  10435   &  10423   \\
 $2^{--}(2D)$  &                            &   10476   &  10449  &  10443   &  10424   \\
 $2^{-+}(2D)$  &                            &   10475   &  10450  &  10445   &  10424   \\
 $3^{--}(2D)$  &                            &   10478   &  10455  &  10449   &  10423   \\
 $1^{--}(3D)$  & 10752.7$\pm$5.9$^{+0.7}_{-1.1}$ &   10742   &  10698  &  10704   &  10731   \\
 $2^{--}(3D)$  &                            &   10744   &  10705  &  10711   &  10733   \\
 $2^{-+}(3D)$  &                            &   10742   &  10706  &  10713   &  10733   \\
 $3^{--}(3D)$  &                            &   10740   &  10711  &  10717   &  10733   \\
 $1^{--}(4D)$  & 10992.9$^{+10.0}_{-3.1}$   &   10987   &  10928  &  10949   &  11013   \\
 $2^{--}(4D)$  &                            &   10986   &  10934  &  10957   &  11016   \\
 $2^{-+}(4D)$  &                            &   10984   &  10935  &  10959   &  11015   \\
 $3^{--}(4D)$  &                            &   10981   &  10939  &  10963   &  11015   \\
\bottomrule[1pt]\bottomrule[1pt]
\end{tabular*}
\end{table}

According to the predicted masses by the RFT model and other methods~\cite{Soni:2017wvy,Godfrey:2015dia,Ebert:2011jc}, the $4D$ $b\bar{b}$ states should have the masses around 10950 MeV. The controversial $\Upsilon(11020)$ state might have a significant $4^3D_1$ component since its mass is quite close to the prediction of the $4^3D_1$ state. Furthermore, the dielectron width of the pure $6S$ $\Upsilon$ state was given about 0.274 KeV~\cite{Badalian:2008ik}, which is about two times larger than the experimental measurement of $\Upsilon(11020)$. This result also indicated that the $S$-$D$ mixing effect should be significant for the $\Upsilon(11020)$ state.

\subsection{High orbital excited states} \label{D}

Up to now, none of the high orbital excited $b\bar{b}$ mesons including $F$-, $G$-, and $H$-wave states have been announced by any experiments. Obviously, it is a challenge for experiments to discover these states. However, the situation may have changed while the SuperKEKB facility has ran last year~\cite{Kou:2018nap}. With the event numbers about 2$\times10^6$ $\Upsilon(2^3D_1)$ states produced at Belle II in future, the observation of $F$-wave $b\bar{b}$ state could be accessible~\cite{Kou:2018nap}.

\begin{table}[htbp]
\caption{The masses of high orbital excited $b\bar{b}$ states (in MeV).
}\label{table7}
\renewcommand\arraystretch{1.2}
\begin{tabular*}{85mm}{c@{\extracolsep{\fill}}ccccc}
\toprule[1pt]\toprule[1pt]
 States        &       Our     &  Ref.~\cite{Godfrey:2015dia}   &  Ref.~\cite{Wang:2018rjg}    &  Ref.~\cite{Ebert:2011jc}   &  Ref.~\cite{Segovia:2016xqb} \\
\midrule[1pt]
 $2^{++}(1F)$  &       10376   &  10350  &  10362   &  10343   &  10315 \\
 $3^{++}(1F)$  &       10391   &  10355  &  10366   &  10346   &  10321 \\
 $3^{+-}(1F)$  &       10391   &  10355  &  10366   &  10347   &  10322 \\
 $4^{++}(1F)$  &       10400   &  10358  &  10369   &  10349   & $\cdots$ \\
 $2^{++}(2F)$  &       10668   &  10615  &  10605   &  10610   &  10569 \\
 $3^{++}(2F)$  &       10670   &  10619  &  10609   &  10614   &  10573 \\
 $3^{+-}(2F)$  &       10668   &  10619  &  10609   &  10615   &  10573 \\
 $4^{++}(2F)$  &       10667   &  10622  &  10612   &  10617   & $\cdots$ \\
 $2^{++}(3F)$  &       10920   &  10850  &  10809   & $\cdots$ &  10782 \\
 $3^{++}(3F)$  &       10918   &  10853  &  10812   & $\cdots$ &  10785 \\
 $3^{+-}(3F)$  &       10916   &  10853  &  10812   & $\cdots$ &  10785 \\
 $4^{++}(3F)$  &       10912   &  10856  &  10815   & $\cdots$ & $\cdots$ \\
 $3^{--}(1G)$  &       10588   &  10529  &  10533   &  10511   &  10506 \\
 $4^{--}(1G)$  &       10592   &  10531  &  10535   &  10512   & $\cdots$  \\
 $4^{-+}(1G)$  &       10591   &  10530  &  10534   &  10513   & $\cdots$  \\
 $5^{--}(1G)$  &       10592   &  10532  &  10536   &  10514   & $\cdots$  \\
 $3^{--}(2G)$  &       10851   &  10769  &  10745   & $\cdots$ &  10712 \\
 $4^{--}(2G)$  &       10848   &  10770  &  10747   & $\cdots$ & $\cdots$ \\
 $4^{-+}(2G)$  &       10846   &  10770  &  10747   & $\cdots$ & $\cdots$ \\
 $5^{--}(2G)$  &       10842   &  10772  &  10748   & $\cdots$ & $\cdots$ \\
 $4^{++}(1H)$  &       10778   &$\cdots$ & $\cdots$ &  10670   & $\cdots$ \\
 $5^{++}(1H)$  &       10776   &$\cdots$ & $\cdots$ &  10671   & $\cdots$ \\
 $5^{+-}(1H)$  &       10774   &$\cdots$ & $\cdots$ &  10671   & $\cdots$ \\
 $6^{++}(1H)$  &       10769   &$\cdots$ & $\cdots$ &  10672   & $\cdots$ \\
\bottomrule[1pt]\bottomrule[1pt]
\end{tabular*}
\end{table}

The masses of the $1F$ $b\bar{b}$ states are predicted in the region around 10400 MeV (see Table \ref{table7}), which is comparable with the results given by the lattice nonrelativistic QCD~\cite{Lewis:2012ir}. The $1G$ $b\bar{b}$ masses are predicted around 10590 MeV which are slightly above the $B\bar{B}$ threshold at 10.56 GeV. Our predicted masses of $1G$ $b\bar{b}$ states seem to be larger than the results given by the quark potential models~\cite{Segovia:2016xqb,Godfrey:2015dia,Wang:2018rjg,Ebert:2011jc}, but very close to the results from the lattice QCD~\cite{Lewis:2012ir}, where the masses of $4^{-+}$ and $4^{--}$ $b\bar{b}$ states were predicted as
\begin{equation}
\begin{split}
\textrm{M}(^1G_4)=&10581\pm17~\textrm{MeV},\\
\textrm{M}(^3G_4)=&10587\pm18~\textrm{MeV}. \label{eq31}
\end{split}
\end{equation}

\section{Further discussions: A comparison of results given by the RFT model and the quark potential model}\label{sec5}

From Tables \ref{table4}$-$\ref{table6}, one may notice that the masses predicted by the quark potential model~\cite{Godfrey:2015dia,Wang:2018rjg,Ebert:2011jc} and the RFT model  are quite similar for these low-lying $b\bar{b}$ states. Since the higher excited bottomonium states have not been found by experiments, there is no criterion from the experimental measurements to distinguish these models. In this section, we give a comparison of the RFT model and the quark potential model.

From a phenomenological point of view, the confinement mechanism for the quarks in a hadron system could be mimicked in two simple ways. In the quark potential model, the confinement mechanism is usually implemented by a long-distance linear potential~\cite{Eichten:1974af,Eichten:1978tg}. Differently, a dynamical flux tube in the RFT model is responsible for the confinement mechanism~\cite{LaCourse:1988cu,Olson:1991tw}. In the following, we compare these two models from three aspects.

First, we may compare the Hamiltonian of the RFT model to the relativized quark potential (RQP) model~\cite{Godfrey:1985xj}, directly. The following spinless Salpeter Hamiltonian of the RQP model
\begin{equation}
H^{\textrm{RQP}}=2\sqrt{p^2+m^2}+\sigma{r}, \label{eq32}
\end{equation}
has been used to calculate the mass spectra of bottomonium states in Ref.~\cite{Godfrey:2015dia} where the Coulomb potential in short range and a mass-renormalized constant $C$ were supplemented. For comparing with the Hamiltonian of Eq. (\ref{eq32}), we rewrite the Hamiltonian of RFT model as
\begin{equation}
\begin{split}
H^{\textrm{RFT}}=&~2\sqrt{p^2+m^2}+\frac{\sigma{r}}{2}\left(\frac{\arcsin{\varepsilon_1}}{\varepsilon_1}+\varepsilon_2\right)\\
&+\frac{\sqrt{p_r^2+m^2}}{16p_t^2}\varepsilon_2\sigma^2r^2\left(\frac{\arcsin{\varepsilon_1}}{\varepsilon_1}-\varepsilon_2\right)^2, \label{eq33}
\end{split}
\end{equation}
which was obtained by an expansion in the string tension $\sigma$~\cite{Brambilla:1992pe}. The parameters $\varepsilon_1$ and $\varepsilon_2$ in Eq.~(\ref{eq33}) were defined as
\begin{equation}
\varepsilon_1=\sqrt{\frac{p_t^2}{p^2+m^2}};~~~~~~~\varepsilon_2=\sqrt{\frac{p_r^2+m^2}{p^2+m^2}}. \label{eq34}
\end{equation}
In Eqs.~(\ref{eq33}) and (\ref{eq34}), $p_r$, $p_t$, and $p$ denote the radial, transverse, and total momentum of the quark which was attached with the flux tube in a meson system. If the mass of the quark $q$ in a $q\bar{q}$ meson tends to infinity (i.e., $m\rightarrow\infty$), we have the limits: $\varepsilon_1\rightarrow{0}$ and $\varepsilon_2\rightarrow{1}$, since the $p_r$, $p_t$ and $p$ are far smaller than the quark mass. In this limit, the flux tube model reduces to a quark model with linear confinement potential~\cite{Allen:2003wz}. But the realistic mass of $b$ quark in the bottomonium system is finite; the contribution of flux tube cannot reduce to a simple static potential~\cite{Buisseret:2007de}.

Secondly, the following formula of excited energy which was obtained by the RFT model (also see Eq.~(\ref{eq18}) in Sec.~\ref{sec2})
\begin{equation}
E^{\textrm{RFT}}_{nL}=\left(\frac{\sigma^2}{2\pi^2m_b}\right)^{1/3}(1.41n+L)^{2/3}, \label{eq35}
\end{equation}
is also different from the result which was given by the quark potential model. In principle, the spinless Salpeter equation or the Schr\"{o}dinger equation with linear confinement potential can hardly be solved analytically. But the Schr\"{o}dinger equation with linear potential, i.e., the nonrelativistic version of Eq.~(\ref{eq32}), can be solved approximately by the perturbation expansion method~\cite{Bose:1975zx}, the Wentzel-Kramers-Brillouin approach~\cite{Quigg:1979vr}, the variational method~\cite{Karl:1994ji}, and the shifted $1/N$ expansion method~\cite{Sukhatme:1983ce}. By comparing with the numerical results in Refs.~\cite{Karl:1994ji,Richard:1992uk}, one may find that precision of the approximate solution obtained by the variational method~\cite{Karl:1994ji} is best for the excited energy of meson systems. According to the results in Ref.~\cite{Karl:1994ji}, we could write the energy formula as
\begin{equation}
E^{\textrm{Var.}}_{nL}=\left(\frac{6.645\sigma^2}{m_b}\right)^{1/3}(1.80n+L+1.40)^{2/3}. \label{eq36}
\end{equation}
This is an approximate formula for the excited energy of low radial $b\bar{b}$ excitations, which could be regarded as the approximate solution of the Schr\"{o}dinger equation with linear confinement potential. Obviously, it is quite different from Eq.~(\ref{eq35}) which was deduced from the RFT model.

Finally, we may directly compare the spin average masses of $b\bar{b}$ states, which were predicted by the RFT model, to the results given by a nonrelativistic constituent quark model~\cite{Segovia:2016xqb}. The concrete results of the corresponding $n^{2S+1}L_J$ multiplet with their differences are listed in Table~\ref{table8}. Obviously, the differences of predicted masses given by two models are smaller than 50 MeV for these low excited $b\bar{b}$ states, including 3$S$, 4$S$, 2$P$, 3$P$, and 1$D$ states. However, the discrepancy of predictions becomes large for the higher excited bottomonium states. Especially for the high orbital excitations, the differences of predicted masses are very remarkable. This interesting result can be naturally explained since the flux tube can carry both angular momentum and energy. In fact, this point is conceptually different from the quark potential models~\cite{LaCourse:1988cu}.

\begin{table}[htbp]
\caption{A comparison of the spin average masses which were predicted by the RFT model and the nonrelativistic constituent quark model~\cite{Segovia:2016xqb} (in MeV).
}\label{table8}
\renewcommand\arraystretch{1.2}
\begin{tabular*}{85mm}{l@{\extracolsep{\fill}}ccccc}
\toprule[1pt]\toprule[1pt]
 $n^{2S+1}L_J$               &   $1S$     &   $2S$       &  $3S$    &  $4S$     &  $5S$     \\

 Our                         &   Input    &  Input       &  10352   &  10634    &  10885    \\
 Ref.~\cite{Segovia:2016xqb} &   9490     &  10009       &  10344   &  10607    &  10818    \\
 $\delta{M}$                 &  $\cdots$  &  $\cdots$    &  8       &  27       &  67        \\
\toprule[1pt]
 $n^{2S+1}L_J$               &   $1P$     &   $2P$       &  $3P$    &  $4P$     &  $5P$      \\

 Our                         &   Input    &   10262      &  10556   &  10815    &  11051      \\
 Ref.~\cite{Segovia:2016xqb} &   9879     &   10240      &  10516   &  10744    &  $\cdots$   \\
 $\delta{M}$                 &   $-$      &   22         &  40      &  71       &  $\cdots$   \\
\toprule[1pt]
 $n^{2S+1}L_J$               &   $1D$     &   $2D$       &  $3D$    &  $4D$     &   $5D$      \\

 Our                         &   10167    &   10475      &  10742   &  10984    &  11209      \\
 Ref.~\cite{Segovia:2016xqb} &   10123    &   10419      &  10658   &  10860    &  $\cdots$      \\
 $\delta{M}$                 &   44       &   56         &  84      &  124      &  $\cdots$   \\
\toprule[1pt]
 $n^{2S+1}L_J$               &   $1F$     &   $2F$       &  $3F$    &  $1G$     &  $2G$        \\

 Our                         &   10391    &   10668      &  10916   &  10591    &  10846       \\
 Ref.~\cite{Segovia:2016xqb} &   10322    &   10573      &  10785   &  10506    &  10712       \\
 $\delta{M}$                 &   69       &   95         &  131     &  85       &  134         \\
\bottomrule[1pt]\bottomrule[1pt]
\end{tabular*}
\end{table}

\section{Conclusion and summary}\label{sec6}

In this work, we derived a Chew-Frautschi like formula which can give an intuitive description of the spin average mass of the heavy quarkonium systems. With the measured masses of $1S$, $2S$, and $1P$ $b\bar{b}$ states, we fixed the three parameters in the Chew-Frautschi like formula, namely, the mass of $b$ quark, string tension $\sigma$, and dimensionless parameter $\lambda$. Then we tested the mass formula by comparing the predicted spin average masses of $3S$, $2P$, and $1D$ states to the experimental results. The comparison implied that the Chew-Frautschi like formula could describe the spin average masses of high excited $b\bar{b}$ states well.

Inspired by a good description of the spin average mass, we further incorporate the spin-dependent interactions which include the OGE 
forces and the longer-ranged inverted spin-orbit term. As shown in the Tables \ref{table4} and \ref{table5}, the measured masses of the $nS$ ($2\leq{n}\leq6$) and $nP$ ($n=1$ and 2) states were well reproduced. The predicted masses of $nD$ and other high bottomonium states in Tables \ref{table6} and \ref{table7} could be tested in future.

In addition, the differences between the RFT model and the quark potential model have also been discussed. To further reveal the role of the flux tube in the RFT model, we also compared the masses predicted by a nonrelativistic constituent quark model~\cite{Segovia:2016xqb} and  the RFT model. We list the main conclusions below.

\begin{enumerate}[(1)]
\item The $\Upsilon(10860)$ could be explained as a predominant $5S$ state since its measured mass is very close to the predictions (see Table \ref{table4}). The $\Upsilon(10580)$ and $\Upsilon(11020)$ cannot be regarded as the pure $4S$ and $6S$ states, respectively, since the predicted masses are much larger than the measurements.

\item The newly discovered $\Upsilon(10750)$ could be regarded as a good candidate of the predominant $3^3D_1$ state since the measured mass is in good agreement with our prediction.

\item The measured masses of $3P$ $b\bar{b}$ states seem to be about 20$-$30 MeV smaller than the theoretical results.

\item Our predicted mass of the $1^3D_2$ $b\bar{b}$ state is consistent with the experimental value. The predicted masses of $1^3D_1$ and $1^3D_3$ states are also comparable with the signals detected by the CLEO~\cite{Bonvicini:2004yj} and $BABAR$~\cite{delAmoSanchez:2010kz} collaborations.
\end{enumerate}

In summary, the bottomonium spectrum has been systematically studied by the RFT model, which could be regarded as an important supplement to the available investigations of the bottomonium spectrum. Since the relativistic color flux tube carries both energy and momentum, the RFT model presents a different dynamics picture for the heavy quarkonia system. The larger predicted masses of the high orbital excited states by the RFT model can be tested by the experiments in future. Combining with our previous work~\cite{Chen:2014nyo}, the RFT model has provided a reasonable scheme to describe the masses of single heavy baryons and heavy quarkonium. We may try to extend the RFT model to analyze the mass spectrum of light meson system in future.


\begin{acknowledgments}
 We thank Professor Xiang Liu for the helpful suggestions and A. Bondar for telling us the discovery history of the $\eta_b(2S)$ state. This research was supported in part by the National Natural Science Foundation of China under Grants No. 11305003 and No. 11975146.
\end{acknowledgments}

\end{document}